\documentstyle[preprint,aps]{revtex}
\begin{document}
\draft

\title{Finite-size scaling of the quasispecies model}

\author{P. R. A. Campos  and J. F. Fontanari }

\address{
Instituto de F\'{\i}sica de S\~ao Carlos  \\ Universidade de S\~ao
Paulo \\ Caixa Postal 369 \\ 13560-970 S\~ao Carlos SP \\ Brazil}

%\date{}

\maketitle

\begin{abstract}

We use finite-size scaling to study
the critical behavior of the quasispecies model of molecular
evolution in the single-sharp-peak replication landscape.
This model exhibits a sharp threshold phenomenon at $Q = Q_c = 1/a$,
where $Q$ is the probability of exact replication of a molecule
of length $L$ and $a$ is the selective advantage of the
master string. We investigate the sharpness of the threshold and
find that its characteristics persist
across a range of $Q$ of order $L^{-1}$ about $Q_c$. Furthermore,
using the data collapsing method we show that the normalized
mean Hamming distance between the master string and the entire population,
as well as  the properly scaled fluctuations around this mean value,  
follow  universal forms in the critical region.
 
\end{abstract}

\pacs{87.10.+e, 64.60.Cn }

%\newpage

Although the so-called  error threshold phenomenon, which limits
the length $L$  of competing  self-reproducing molecules, 
is acknowledged  as one of the main outcomes of Eigen's 
quasispecies model \cite{Eigen,reviews}, the full characterization of 
the error threshold transition for finite $L$
has not been satisfactorily carried out yet. 
In fact, similarly to the
definition of the critical temperature for finite lattices,
there is no generally accepted definition
of the term error threshold for finite $L$ \cite{Wiehe}. 
Nevertheless, the study of 
the systematic deviations from the infinite length limit
behavior introduced by the finite-size effects, 
besides being practically  independent of the 
definition adopted, gives valuable information on the
behavior of the relevant macroscopic quantities near the
critical region \cite{Barber,Binder}.

In the quasispecies model, a molecule is represented by a string of 
$L$ digits 
$\vec{s} = \left (s_1, s_2, \ldots, s_L \right )$, with
the variables $s_\alpha$ allowed to take on $\kappa$ different
values, each  representing a different type of monomer used to build 
the molecule. For sake of simplicity,
in this work  we will consider only binary strings, i.e., 
$s_\alpha = 0,1$. 
The concentrations $x_i$ of molecules of 
type $i =1, 2,  \ldots, 2^L $ evolve in time according to 
the following differential equations \cite{Eigen,reviews}
\begin{equation}\label{ODE}
\frac{dx_i}{dt} = \sum_j W_{ij} x_j - \left [ D_i + \Phi \left ( t
\right ) \right ] x_i \;  ,
\end{equation}
where the constants $D_i$ stand for the death probability of 
molecules of type $i$, and $\Phi (t)$ is a dilution flux 
that keeps the total concentration constant. This flux 
introduces a nonlinearity in  (\ref{ODE}), and is determined
by the condition $ \sum_i dx_i /dt = 0$. More pointedly,
assuming $D_i = 0$ for all $i$ and $\sum_i x_i = 1$ yields
\begin{equation}\label{flux}
\Phi = \sum_{i,j} W_{ij} x_j .
\end{equation}
The  elements of the replication matrix $ {\bf W} $ are given by 
\begin{equation}
W_{ii} = A_i \,  q^L
\end{equation}
and 
\begin{equation}
W_{ij} = A_i \,
 q^{L - d \left ( i,j \right )} 
 \left ( 1 - q \right )^{d \left ( i,j \right )} ~~~~i \neq j ,
\end{equation}
where $A_i$ is the replication rate or fitness of molecules of type $i$,
and $d \left ( i,j \right )$ is the Hamming distance
between strings  $i$ and $j$. Here $0 \leq q \leq 1$ is
the single-digit replication accuracy, which is assumed to be the same 
for all digits. 

In this work we will consider the simplest and probably most
studied replication landscape, namely the single-sharp-peak replication 
landscape, in which we 
ascribe the replication rate $a > 1$ to  the so-called master string
$\left (0, 0, \ldots, 0 \right )$, and the replication rate $1$ to  the
remaining strings. 
In this context, the parameter $a$ is termed selective advantage
of the master string.
As the replication accuracy $q$ decreases, two distinct
regimes are observed in the population composition: the 
{\em quasispecies} regime characterized
by the master string and its close neighbors, and the 
{\em uniform} regime where the $2^L$ strings appear in the
same proportion. 
The transition between these regimes occurs at the error threshold
$q_c$. To study this transition
for large $L$ it is more convenient to introduce the probability
of exact replication of an entire  string, namely
\begin{equation}\label{Q}
      Q = q^L ,
\end{equation}
so that  for $L \rightarrow \infty$  the transition  occurs at
\cite{Eigen,reviews} 
\begin{equation}\label{Q_c}
      Q_c = \frac{1}{a} .
\end{equation}

Although there is a consensus that a thermodynamic order-disorder 
phase transition occurs in the limit $L \rightarrow \infty$ only
\cite{Lethausser,Tarazona,Gal}, there is some disagreement on the 
order of the transition. On the one hand, the mapping of the 
steady-state solution of the chemical kinetic equations (\ref{ODE})
into the surface properties of a semi-infinite two-dimensional
lattice system in thermodynamic equilibrium indicates that 
the relevant order parameter, namely, the mean normalized Hamming distance 
$d$
between the master string and the entire population, vanishes
continuously at $Q_c$ \cite{Tarazona}. However, due to the enormous
difficulty of solving the self-consistent equations that describe
the equilibrium surface properties that analysis was restricted
to $L = 20$ \cite{Tarazona}.
On the other hand, a thorough  investigation of
an alternative mapping of equations (\ref{ODE})  into a  problem of
directed polymers in a random medium 
indicates that the concentration of master strings 
presents a discontinuity at $Q=Q_c$ \cite{Gal}. Since
this mapping allows for the exact solution of the quasispecies
model in the single-sharp-peak replication landscape for generic
lengths $L$, that result implies that  the transition for
$L \rightarrow \infty$ is definitely of first order \cite{Gal}.  

The aim of this work is to investigate the finite-size effects
near the error threshold transition.
Of particular interest is the determination of the sharpness of 
the threshold, namely,  the range of $Q$ about $Q_c$ where
the threshold characteristics persist. As we expect
that the size of this region shrinks to zero like $L^{-1/\nu}$
as $L \rightarrow \infty$, our  goal is to estimate the
value of the exponent $\nu \geq 0$ using finite-size scaling
or, more precisely, the data collapsing method \cite{Binder}.
Our approach is   
in the same spirit of the finite-size scaling 
of combinatoric problems \cite{Kirk}, for which there is no geometric 
criterion for defining a quantity analogous to the correlation length
$\xi$,
and so the success of the method in accounting for the size
dependence of the order parameters cannot be attributed to 
the divergence of $\xi$ and the consequent  onset of a second order 
phase transition. 
In fact, instead of attempting to map the chemical kinetic equations
(\ref{ODE})
into a equilibrium statistical mechanics problem, we resort to a
simpler and more direct approach, namely, the  exact numerical 
solution of those equations in the steady-state regime for  molecule
lengths up to $L = 150$. 

As pointed out by Swetina and Schuster \cite{Swetina},
for the single-sharp-peak replication landscape
the $2^L$ molecular concentrations $x_i$ can be grouped into
$L +1$ distinct classes according to their Hamming distances to
the master string. This procedure allows the  description of
the chemical kinetics by the following $L + 1$ coupled first-order
differential equations \cite{Swetina}
\begin{equation}\label{sw}
\frac{dY_P}{dt} = \sum_{R=0}^{L} M_{PR} \, Y_R  + \left ( a -1 \right )
 Y_0 M_{P 0}
- Y_P \left [ 1 + Y_0 \left ( a - 1 \right ) \right ] ,  
\end{equation}
where $Y_P$ denotes the concentrations 
of molecules in class $P = 0,\ldots, L$. Clearly,  $\sum_P Y_P = 1$.
Here $M_{PR}$ stands for the
probability of mutation from a molecule of type $R$ to a
molecule of type $P$ and is given by 
\begin{equation}\label{M}
M_{P R} = \sum_{I = I_l}^{I_u} \left ( \! \! \begin{array}{c} R \\ I 
\end{array}
\! \! \right ) \, \left ( \! \! \begin{array}{c} L - R \\ P - I
\end{array} \! \! \right ) \, q^{L - P - R + 2I} \,
\left ( 1 - q \right )^{P + R - 2I} ,
\end{equation}
where $I_l = \mbox{max} \left ( 0,P+R-L \right)$ and
$I_u = \mbox{min} \left ( P,R  \right)$.

The procedure to obtain the steady-state solution $dY_P/dt = 0$ of
Eqs.\ (\ref{sw}) is straightforward. The steady-state concentrations 
$Y_P$ for $P=0,\ldots,L$  can be  easily found 
by solving  by iterations the following set of equations
\begin{equation}\label{le}
Y_P = \frac{ \sum_{R=0}^{L} M_{PR} \, Y_R  + \left ( a -1 \right )
 Y_0 M_{P 0} }{ 1 + Y_0 \left ( a - 1 \right ) } .  
\end{equation}
Interestingly, the iteration of these equations is identical
to the dynamics of a recently  proposed population genetics model 
based on the neglect of the linkage disequilibrium at the population level
\cite{Alves}. 
 
The relevant quantities to describe the structure of the population
are the normalized mean Hamming distance
between the master string and the whole population, defined by
\begin{equation}
d = \frac{1}{L} \sum_{P=0}^L  P \, Y_P ,
\end{equation}
and the average of the squared deviations  around $d$, 
\begin{equation}
\sigma^2 =  L^2 ~\sum_{P=0}^L 
\left (  \frac{P}{L} - d \right )^2 \,Y_P.
\end{equation}
Clearly, $d$ and $\sigma^2$ are the analogous to the magnetization 
and susceptibility in a system of Ising spins.
In Figs.\ \ref{f1} and \ref{f2} we present $d$ and  $\sigma^2$,
respectively, as functions of the properly normalized 
probability of exact replication $Q/Q_c$. As expected, the results 
of Fig.\ \ref{f1} show the sharpening of the transition with increasing
$L$.  Furthermore, all curves intersect at an unique point, namely,
the critical point $Q = Q_c$. This somewhat unexpected result
has proved very useful to locate the threshold  
in the case that its location  is not known apriori \cite{Kirk}. 
The curves shown in Fig.\ \ref{f2}  indicate that the height of the peak 
of $\sigma^2$, denoted by $\sigma^2_{\mbox{\small max}}$, 
increases with increasing $L$ like  $L^{\gamma/\nu}$. 
As illustrated in the inset, the ratio $\gamma/\nu$ is given by the 
slope of the straight line fitting the data points in a plot of 
$\ln \sigma^2_{\mbox{\small max}} $ versus $\ln L$.
The result 
$\gamma/\nu = 1.96 $ is in  good agreement with the analytical
prediction that the rms amplitude of a quasispecies around
the master string ($\sqrt{\sigma^2}$) is found to diverge 
algebrically with the exponent 1 as $Q \rightarrow Q_c$ \cite{Gal}.

The exponent $1/\nu$ is estimated using the standard
data collapsing method \cite{Binder} as illustrated in Figs.\ \ref{f3} 
and \ref{f4}, where we plot $d$ and $ L^{-\gamma/\nu} \sigma^2$, 
respectively, versus  $L^{1/\nu} \epsilon$. Here
$\epsilon = \left ( Q - Q_c \right )/Q_c$ is the
reduced probability of exact replication.
The collapse of the curves for different $L$ was
achieved with the exponents $1/\nu = 1$ and $\gamma/\nu = 1.958$
regardless of the value of the selective advantage parameter $a$,
indicating then the universal character of these exponents.
However, as shown in these figures, the universal forms 
(i.e., scaling functions) 
followed by  the properly scaled order parameters in the critical region
depend on $a$. 
Since $ \nu = 1$, we note that the
characteristics of the threshold transition persist
across a range of $Q$ of order $L^{-1}$ about $Q_c = 1/a$.

As in the case of combinatoric problems \cite{Kirk}, it is surprising
that finite-size scaling is so effective to  characterize the 
error threshold transition of the quasispecies model in the 
single-sharp-peak 
replication landscape, which is known to be of first order \cite{Gal}. 
The  collapse of the data for different $L$
into a single, universal  curve presented in Figs.\ \ref{f3} and \ref{f4},
however, is an incontestable evidence of the usefulness of the
finite-size scaling method to investigate threshold phenomena.
In fact, the existence of the universal forms presented in those
figures together with the characterization of the sharpness
of the error threshold are the main results of this paper.

\acknowledgments
The work of JFF was supported in part by Conselho Nacional de 
Desenvolvimento Cient\'{\i}fico e Tecnol\'ogico (CNPq). PRAC
holds a FAPESP fellowship.

\begin{figure}
\caption{Normalized mean Hamming distance between  the
master string and the whole population $d$ as
a function of the normalized probability of exact replication $Q/Q_c$  
for $a = 10$, and  $L = 70$ ($\Box$),
$100$ ($\bigcirc$), $120$ ($\triangle$) and $150$ ($\times$).
\label{f1}}
\end{figure}

\begin{figure}
\caption{Standard deviation $\sigma^2$ as
a function of the normalized probability of exact replication $Q/Q_c$.
The inset illustrates the procedure used to estimate the
ratio $\gamma/\nu$.
The parameters and convention are the same as for Fig.\ \ref{f1}.
  \label{f2}}
\end{figure}

\begin{figure}
\caption{Normalized mean Hamming distance as
a function of the scaled reduced probability of exact replication.
The parameters  are $1/\nu = 1$  and (from  bottom to top)
$a=10$, $20$ and  $50$.
The convention is the same as for Fig.\ \ref{f1}.
  \label{f3}}
\end{figure}

\begin{figure}
\caption{Scaled standard deviation  as
a function of the scaled reduced probability of exact replication.
The parameters  are $1/\nu = 1$, $\gamma/\nu = 1.958$,  and 
(from top to bottom at the peak location) $a=10$, $20$ and  $50$.
The convention is the same as for Fig.\ \ref{f1}.
  \label{f4}}
\end{figure}


\begin{references}

\bibitem{Eigen} M. Eigen, Naturwissenchaften {\bf 58}, 465 (1971).

\bibitem{reviews} M. Eigen, J. McCaskill and P. Schuster, 
Adv.\ Chem.\ Phys.\ {\bf 75}, 149 (1989).

\bibitem{Wiehe} T. Wiehe, Genet.\ Res.\ Camb.\ {\bf 69}, 127
(1997).

\bibitem{Barber} M. N. Barber, in
{\em  Phase Transitions and Critical Phenomena}, edited by 
C. Domb and J. L.  Lebowitz (Academic Press, London, 1983),
Vol. 8.

\bibitem{Binder} K. Binder, J.\ Comp.\ Phys.\ {\bf 59}, 1 (1985).

\bibitem{Lethausser} I. Leuth\"{a}usser, J.\ Chem.\ Phys.\ {\bf 84}, 
1884 (1986); J.\ Stat.\ Phys.\ {\bf 48}, 343 (1987).

\bibitem{Tarazona}P. Tarazona, Phys.\ Rev.\ A {\bf 45}, 6038 (1992).

\bibitem{Gal} S. Galluccio, Phys.\ Rev.\ E {\bf 56}, 4526 (1997).

\bibitem{Kirk} S. Kirkpatrick and B. Selman, Science {\bf 264}, 1297
(1994).

\bibitem{Swetina} J. Swetina and P. Schuster, Biophys.\ Chem.\
{\bf 16}, 329 (1982).

\bibitem{Alves} D. Alves and J. F. Fontanari, Phys.\ Rev.\ E
at press.



\end{references}
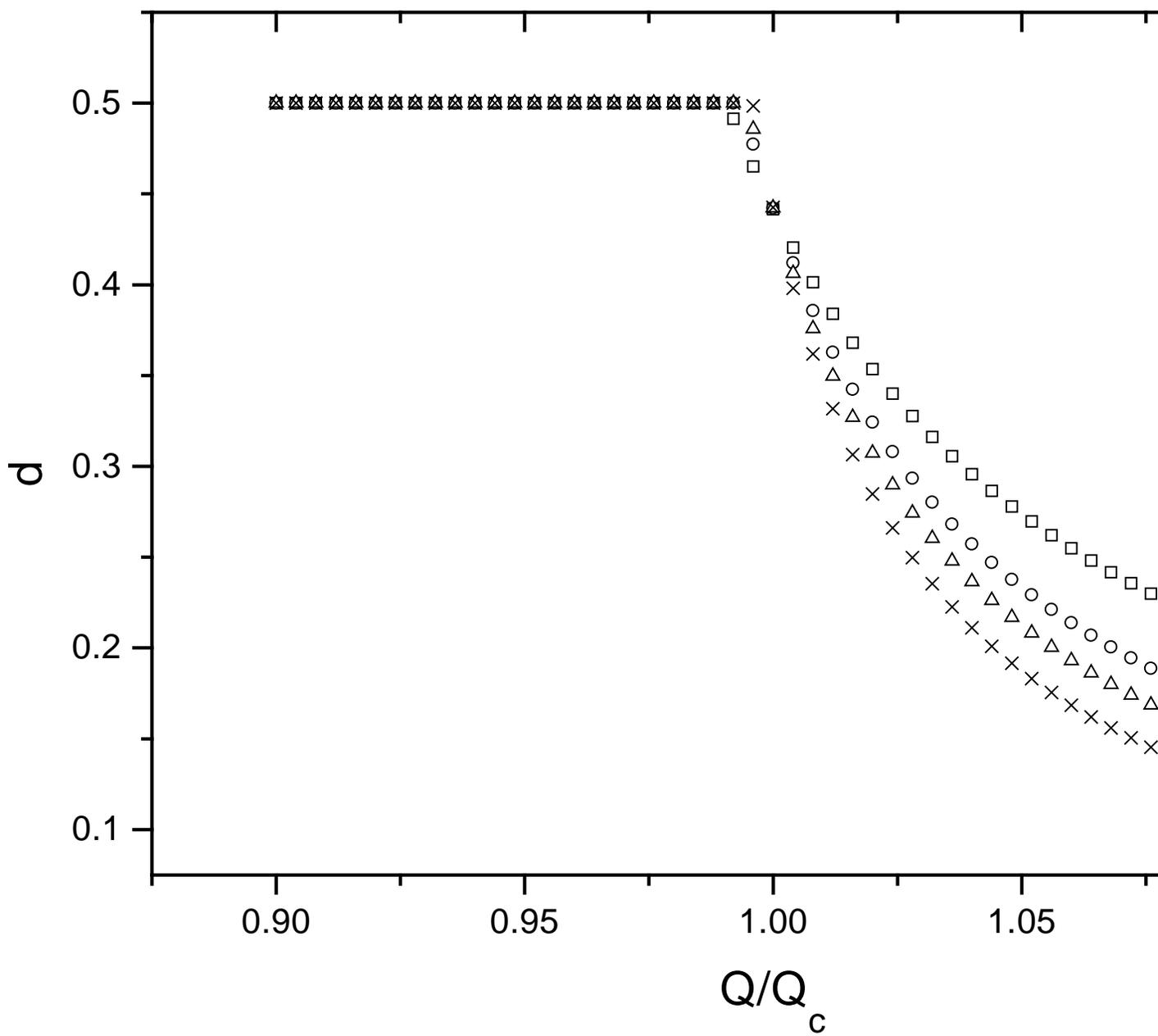
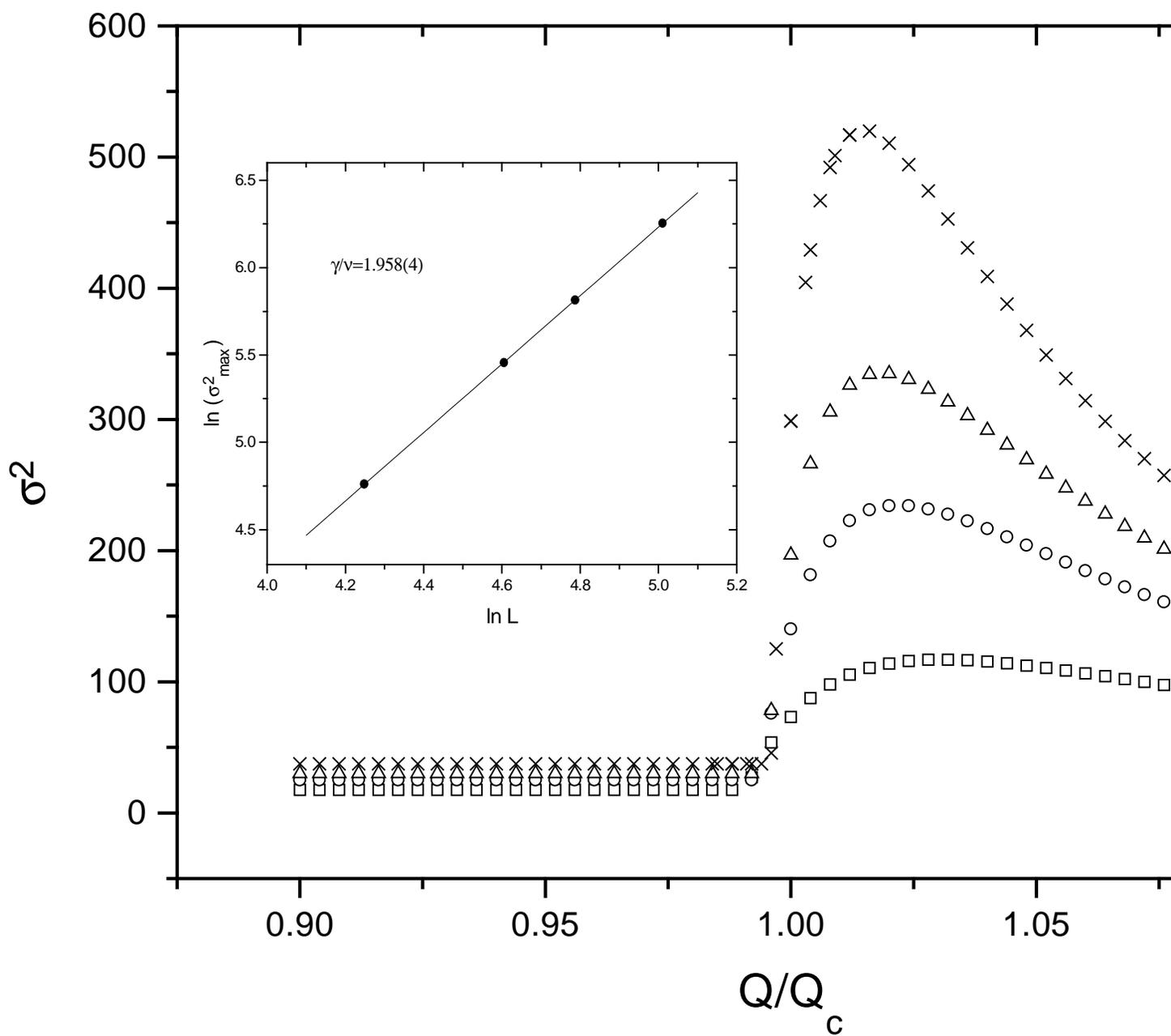
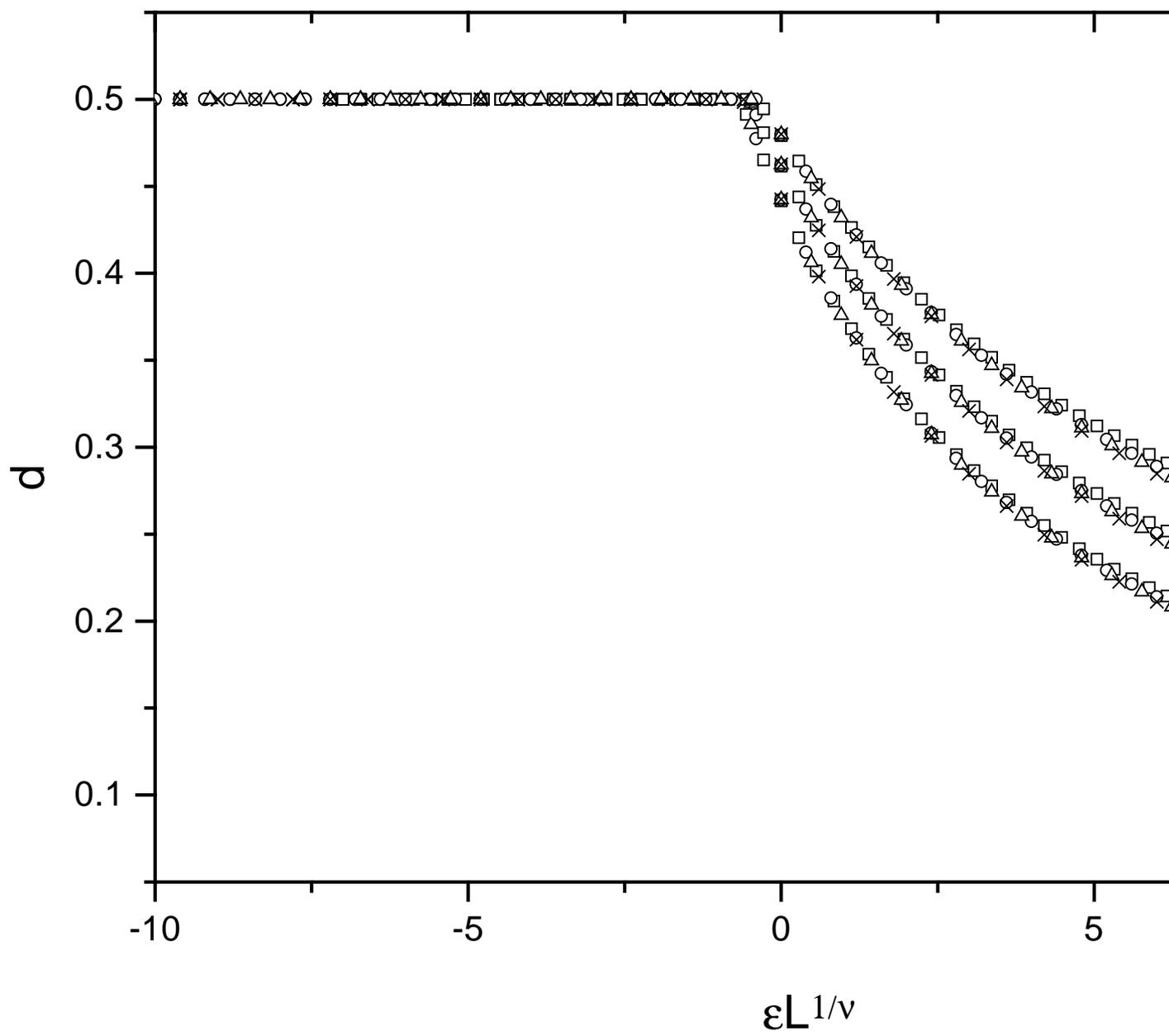
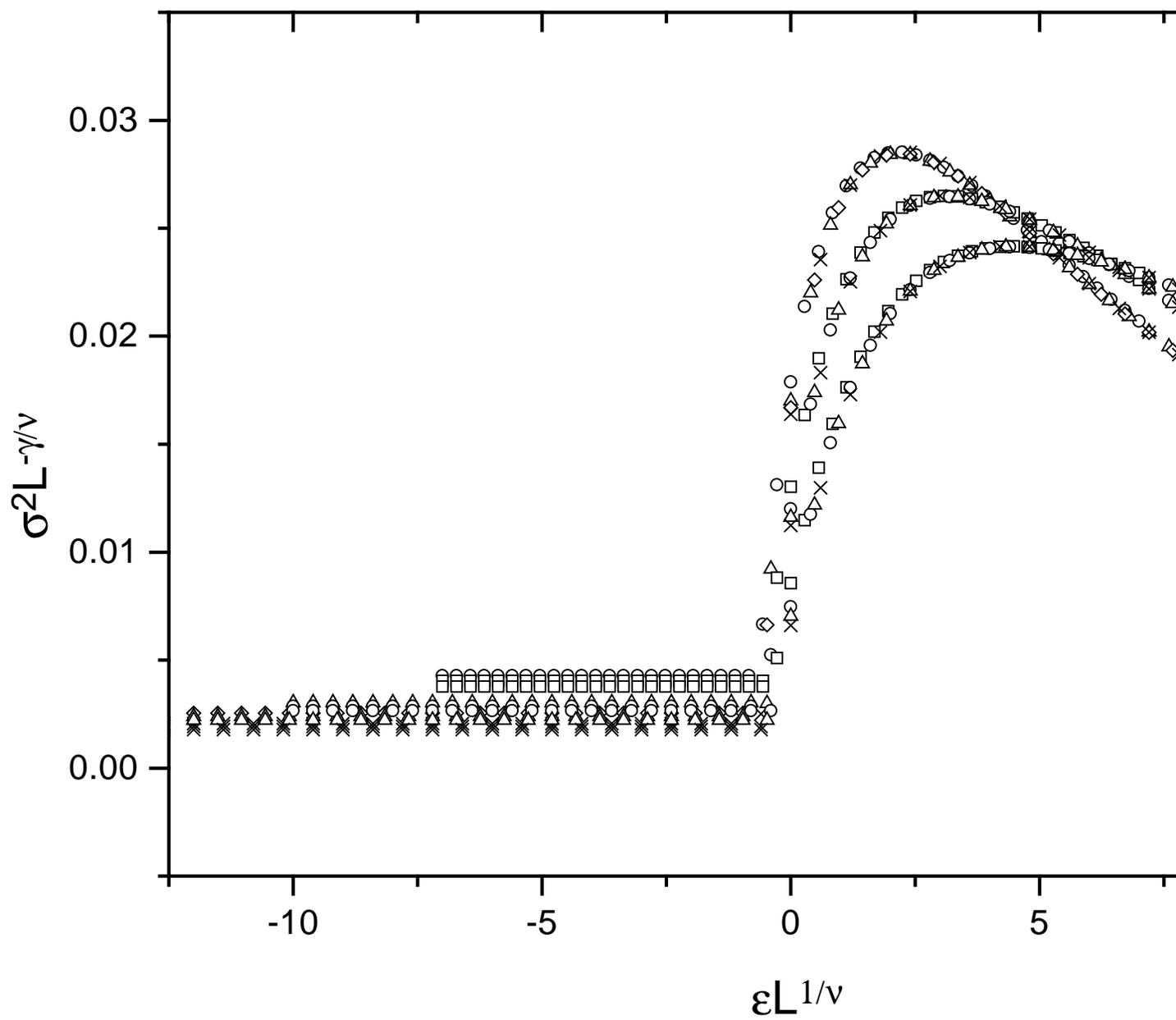
\end{document}